\documentclass[11pt]{article}

\usepackage{amsmath}
\usepackage{amsthm}
\usepackage{amssymb}
\usepackage{comment}
\usepackage{enumerate}
\usepackage{url}

\newcommand{\uinvnorm}{|\kern-1pt|\kern-1pt|}

\newcommand{\supp}{\operatorname{supp}}

\newcommand{\sgn}{\operatorname{sgn}}

\newcommand{\rk}{\operatorname{rank}}

\newcommand{\vcdim}{\operatorname{VC-dim}}

\newtheorem{thm}{Theorem}
\newtheorem*{thm*}{Theorem}
\newtheorem{cor}[thm]{Corollary}
\newtheorem{con}[thm]{Conjecture}
\newtheorem{lem}[thm]{Lemma}
\newtheorem*{lem*}{Lemma}
\newtheorem{prop}[thm]{Proposition}

\newtheorem{obs}[thm]{Observation}

\newcommand{\be}{\begin{equation}}
\newcommand{\ee}{\end{equation}}
\newcommand{\bea}{\begin{eqnarray}}
\newcommand{\eea}{\end{eqnarray}}
\newcommand{\bes}{\begin{equation*}}
\newcommand{\ees}{\end{equation*}}
\newcommand{\beas}{\begin{eqnarray*}}
\newcommand{\eeas}{\end{eqnarray*}}

\newcommand{\R}{\mathbb{R}}

\newcommand{\Z}{\mathbb{Z}}

\newcommand{\F}{\mathbb{F}}
\newcommand{\ip}[2]{\langle #1,#2 \rangle}

\title{On the communication complexity of XOR functions}

\author{Ashley Montanaro\footnote{Department of Computer Science, University of Bristol, Woodland Road, Bristol, BS8 1UB, UK; {\tt montanar@cs.bris.ac.uk}.} \ and Tobias J.\ Osborne\footnote{Department of Mathematics, Royal Holloway, University of London, Egham, TW20 0EX, UK; {\tt tobias.osborne@rhul.ac.uk}.}}

\date{\today}

\begin{document}
	
\maketitle

\begin{abstract}
An XOR function is a function of the form $g(x,y) = f(x \oplus y)$, for some boolean function $f$ on $n$ bits. We study the quantum and classical communication complexity of XOR functions. In the case of exact protocols, we completely characterise one-way communication complexity for all $f$. We also show that, when $f$ is monotone, $g$'s quantum and classical complexities are quadratically related, and that when $f$ is a linear threshold function, $g$'s quantum complexity is $\Theta(n)$. More generally, we make a structural conjecture about the Fourier spectra of boolean functions which, if true, would imply that the quantum and classical exact communication complexities of all XOR functions are asymptotically equivalent. We give two randomised classical protocols for general XOR functions which are efficient for certain functions, and a third protocol for linear threshold functions with high margin. These protocols operate in the symmetric message passing model with shared randomness.
\end{abstract}


\section{Introduction}

The communication complexity model was introduced by Yao in 1979 \cite{yao79}. In its most basic form, the model considers two separated parties (Alice and Bob), who attempt to compute some function $f(x,y)$ of their joint inputs $x$, $y$, while using the minimum amount of communication. They may be required to compute $f$ exactly (the {\em deterministic} model), or may be allowed some constant probability of error (the {\em bounded-error} model). The considerable theoretical interest of this simple model, as well as its practical applications, have motivated its intensive study over the last thirty years.

More recently, the model of {\em quantum} communication complexity was introduced \cite{yao93,kremer95}. In this model, Alice and Bob are allowed to send and receive qubits (quantum bits), with the goal being to reduce the amount of communication required. It has recently been shown that, when the function $f(x,y)$ to be computed is partial (there is some promise on the inputs $x$, $y$), there can be an exponential separation between quantum and classical communication complexity \cite{raz99,gavinsky07}. No separation beyond quadratic is known for {\em total} functions, and it is conjectured that the quantum and classical communication complexities of total functions are in fact polynomially related. However, this conjecture has resisted proof in both the exact and bounded-error models.

A natural way to make progress on the conjecture is to attempt to prove it for restricted types of function. The class of functions $g(x,y) = f(x \wedge y)$, where $f$ is a boolean function, has received particular attention. These functions seem to have first been considered by Buhrman and de Wolf \cite{buhrman01a}, who showed that deterministic quantum and classical communication complexities are asymptotically equivalent for all {\em symmetric} functions $f$ ($f$ is said to be symmetric if $f(z)$ depends only on $|z|$, the Hamming weight of $z$). They also showed that these communication complexity measures are polynomially related if $f$ is a {\em monotone} function ($f$ is said to be monotone if $f(x \vee y) \ge \max\{f(x),f(y)\}$ for all $x$, $y$). It was several more years before Razborov proved, in a fundamental paper \cite{razborov03}, that the bounded-error quantum and classical communication complexities of symmetric functions in this model are polynomially related; see \cite{sherstov08} for a recent alternative proof. In other recent work, Sherstov has shown that the conjecture does indeed hold if one is required to compute both $f(x \vee y)$ and $f(x \wedge y)$ \cite{sherstov09}, and Shi and Zhu have proven lower bounds in a model with a more general notion of composition of functions \cite{shi08}.

This paper is concerned with another natural class of functions, where Alice and Bob each receive an $n$-bit string, and the function they need to compute is defined as $g(x,y) = f(x \oplus y)$ for some boolean function $f$. These functions were recently studied by Shi and Zhang \cite{shi09}, who refer to them as ``XOR functions''. Shi and Zhang essentially determined the quantum and classical communication complexity of all XOR functions where $f$ is symmetric, up to polylogarithmic factors\footnote{Some general quantum lower bounds, which are tight for some XOR functions, had previously been obtained by Buhrman and de Wolf \cite{buhrman01a}, and also Klauck \cite{klauck07}.}. In particular, using Fourier-analytic techniques, they showed that the exact quantum communication complexity of all symmetric XOR functions (excluding a few trivial special cases) is $\Omega(n)$. Bounded-error communication complexity is dealt with via a reduction to the previous result of Razborov \cite{razborov03}. The special case where $f$ is a threshold function ($f(z) = 1 \Leftrightarrow |x|\ge t$ for some $t$) had been considered previously by Huang et al \cite{huang06}.

In the present work, we consider more general classes of XOR function, for which we find new quantum lower bounds and classical upper bounds on communication complexity. As well as monotone functions, another class of function in which we will be interested is {\em linear threshold functions}. $f:\{0,1\}^n \rightarrow \{0,1\}$ is said to be a linear threshold function (LTF) if
%
%
\be \label{eqn:ltf} f(x) = \left\{ 
\begin{aligned}
 0 &\;\;\mbox{ if } \sum_{i=1}^n w_i x_i \le \theta \\
 1 &\;\;\mbox{ if } \sum_{i=1}^n w_i x_i > \theta,
\end{aligned}
\right.
\ee
where $w_i, \theta \in \R$. The set $\{w_i\}$ are known as the {\em weights} of $f$, and $\theta$ is called the {\em threshold} of $f$. These functions have been much studied in the computer science literature and elsewhere; see \cite{saks93} for a review.

In the case of the model of communication complexity studied here, LTFs are a particularly natural class to consider, for the following reason. Imagine that Alice and Bob each have a document, and they wish to determine whether their documents differ significantly. In practice, differing at one position may be more significant than differing at another (consider a bioinformatics application where mutations are more likely at particular points on a chromosome). This can be modelled by the task of determining whether a weighted sum of differences between bits held by Alice and bits held by Bob is above a threshold, which is exactly the problem of computing an XOR function defined by an LTF.

The main results we obtain are as follows. First, we completely characterise the deterministic quantum and classical {\em one-way} communication complexity of XOR functions $g(x,y) = f(x \oplus y)$ in terms of an algebraic property of $f$, its Fourier dimension \cite{gopalan09}. We observe a relationship between deterministic two-way communication complexity and the {\em parity decision tree} model introduced in the context of computational learning theory by Kushilevitz and Mansour \cite{kushilevitz91}, and make a structural conjecture about the Fourier spectra of boolean functions which, if true, would imply that the quantum and classical deterministic two-way communication complexity of all XOR functions are asymptotically equivalent.

Turning to probabilistic communication complexity, we first show that one-way protocols cannot be efficient for all XOR functions: indeed, one-way quantum communication complexity can be exponentially larger than two-way classical communication complexity. On the other hand, there are randomised classical protocols in the more restrictive simultaneous message passing (SMP) model with shared randomness\footnote{See Section \ref{sec:prelims} for the definition of this and other terms in this introduction.}, which are efficient for particular XOR functions. Using a previous result of Grolmusz \cite{grolmusz97}, one can give an efficient protocol for those functions $g(x,y)=f(x \oplus y)$ where $f$ has a very low spectral norm ($f$'s Fourier spectrum is ``narrow''). We give a new protocol that is efficient for functions where $f$ is close to a parity function ($f$'s Fourier spectrum is ``tall''), and in particular for functions where $f$ takes the value 1 on a small number of inputs.

Specialising to particular types of XOR function, we first show that the deterministic quantum and classical two-way communication complexities of all {\em monotone} XOR functions are quadratically related. Specialising further, we show that the deterministic two-way communication complexity of all XOR functions where $f$ is an LTF depending on $n$ bits is $\Theta(n)$. Finally, we give a randomised communication protocol for computing LTFs in the SMP model with shared randomness, which is efficient provided that the margin of the LTF in question is high. The protocol generalises previous results \cite{yao03,huang06} on computing threshold functions.

These results are all given more formally in Section \ref{sec:results} below. In order to state them, we will first require some definitions.


\subsection{Preliminaries}
\label{sec:prelims}


\subsubsection{Query complexity and boolean functions}

We will use a number of mostly standard notions from the field of query complexity (see the review \cite{buhrman02} for further details). Let $f:\{0,1\}^n \rightarrow \{0,1\}$ be a function of $n$ bits. The deterministic decision tree complexity of $f$, written $D(f)$, is the minimal number of queries to the input variables $(x_1,\dots,x_n)$ that are necessary to evaluate $f(x)$ with certainty, for any input $x$. A somewhat less familiar complexity measure is the {\em parity decision tree} complexity $D^\oplus(f)$. This is the minimum number of queries necessary to compute $f(x)$ with certainty on any input $x$, where each query the algorithm makes is the parity of any subset of the $n$ bits of $f$'s input. Note that $D^\oplus(f)$ can be considerably smaller than $D(f)$; a trivial example is given by taking $f$ to be the parity function on $n$ bits, where $D(f) = n$, but $D^\oplus(f)=1$. This model was previously studied by Kushilevitz and Mansour \cite{kushilevitz91}, who showed that functions with low parity decision tree complexity can be learnt efficiently from membership queries.

A {\em boolean function} is a function on the boolean cube $\{0,1\}^n$ that takes one of at most two values on all inputs. When studying the query or communication complexity of boolean functions, we are free to relabel these values, as of course this choice makes no difference to the complexity. In particular, we say that both $f:\{0,1\}^n \rightarrow \{0,1\}$ and $f':\{0,1\}^n \rightarrow \{1,-1\}$ are boolean functions. Any boolean function $f:\{0,1\}^n \rightarrow \{0,1\}$ can be written uniquely as a multilinear polynomial in $n$ variables over $\F_2$; $\deg_2(f)$ denotes the degree of this polynomial. The {\em sensitivity} of a boolean function $f$, written $s(f)$, is defined as the maximum, over all bit strings $x$, of the number of neighbours $y$ of $x$ such that $f(y) \neq f(x)$. 


\subsubsection{Communication complexity}

We study several standard models of communication complexity (see the book \cite{kushilevitz97} for further details). In all models, two parties (Alice and Bob) each receive $n$-bit strings $x$, $y$ (respectively), and share a string of public random bits. Their goal is to compute some boolean function $g(x,y)$ using the minimum amount of communication. The matrix $M_{xy} = g(x,y)$ is known as the communication matrix of $g$.

The communication complexity of $g$ in a given model is the total number of bits that are required to be transmitted to compute $g$. In the two-way model of communication complexity, Alice and Bob take it in turns to send bits to each other; we will assume that Alice speaks first and Bob has to output $g(x,y)$. Define $D^{cc}(g)$ to be the mininum total number of bits required to be transmitted for any classical deterministic protocol to compute $g(x,y)$ on any input. Similarly, let $R_2^{cc}(g)$ denote the number of bits required in the worst case for any classical randomised protocol to compute $g(x,y)$ with success probability at least $2/3$ on every input (the ``2'' refers to 2-sided error).

There are quantum generalisations of these models, in which Alice and Bob are allowed to send and receive qubits (quantum bits) \cite{yao93,kremer95}. We also allow them to share an arbitrary prior entangled quantum state. The total number of qubits required to be transmitted between Alice and Bob for them to compute $g$ exactly and with bounded error will be denoted by $Q_E^{cc}(g)$ and $Q_2^{cc}(g)$, respectively. See \cite{dewolf02} for a good introduction to quantum communication complexity. 

Two more restricted scenarios we consider are the one-way and simultaneous message passing (SMP) models. In the one-way model, Alice sends a single message to Bob, who must then use this message and his own input to evaluate $g(x,y)$. The bounded-error classical and quantum complexities in this model will be denoted by $R_2^1(g)$, $Q_2^1(g)$, respectively. A more restricted setting still is the SMP model. Here, Alice and Bob each send a single message to a referee, who performs some computation on the messages and outputs $g(x,y)$. The randomised communication complexity of $g$ in this model is denoted by $R_2^{\|,pub}(g)$; note that, in this paper, we assume that Alice and Bob are still allowed to share public randomness, which the referee can also see.


\subsubsection{Fourier analysis}

We will make heavy use of some basic ideas from the field of Fourier analysis on the group $\Z_2^n$ (see \cite{odonnell07} or \cite{dewolf08} for excellent introductions to this area). Let $[n]$ denote the set $\{1,\dots,n\}$. Then for any positive integer $n$, the set of $2^n$ parity functions $\chi_S:\{0,1\}^n \rightarrow \{1,-1\}$, $\chi_S(x) = (-1)^{\sum_{i \in S} x_i}$, which are indexed by subsets of $[n]$, are known as the characters of the group $\Z_2^n$. Let $f:\{0,1\}^n \rightarrow \R$ be a function on the boolean cube. Then the {\em Fourier coefficients} of $f$ are the set of coefficients, indexed by subsets $S \subseteq [n]$,
\[ \hat{f}(S) = \frac{1}{2^n} \sum_{x \in \{0,1\}^n} \chi_S(x) f(x). \]
The $p$-norms of $f$ on the Fourier side are defined as
\[ \|\hat{f}\|_p = \left( \sum_{S \subseteq [n]} |\hat{f}(S)|^p \right)^{1/p}, \]
with the special cases $\|\hat{f}\|_0 = |\supp \hat{f}|$ (where $\supp \hat{f}$ denotes the set $\{S: \hat{f}(S) \neq 0\}$), $\|\hat{f}\|_\infty = \max_S |\hat{f}(S)|$; of course, the former is not actually a norm. The norm $\|\hat{f}\|_1$ is known as the {\em spectral norm} of $f$. Parseval's equality states that
\[ \|\hat{f}\|_2^2 = \frac{1}{2^n}\sum_{x \in \{0,1\}^n} f(x)^2. \]
We frequently identify $n$-bit strings with their corresponding subsets of $[n]$ (that is, if $x$ is an $n$-bit string,  and $S$ is the subset of $[n]$ whose characteristic vector is $x$, then $\hat{f}(x)=\hat{f}(S)$). The notation $\hat{f}^{\Delta T}$ denotes the function whose Fourier coefficients are all shifted by $T$: $(\hat{f}^{\Delta T})(S) = \hat{f}(S \Delta T)$, with $S \Delta T$ denoting the symmetric difference of the sets $S$ and $T$. Similarly define $f^{\oplus y}(x) = f(x \oplus y)$. One can easily verify that $\chi_{S \Delta T}(x) = \chi_S(x) \chi_T(x)$ for any $S$, $T$, and similarly $\chi_S(x \oplus y) = \chi_S(x) \chi_S(y)$. The Fourier dimensionality of $f$, $\dim f$, is the smallest $k$ such that the Fourier spectrum of $f$ lies in a $k$-dimensional subspace of $\{0,1\}^n$. Finally, note that when we consider the Fourier transform of a boolean function $f$, we will always assume that $f$ is given in the form $f:\{0,1\}^n \rightarrow \{1,-1\}$.


\subsubsection{Linear threshold functions}
\label{sec:ltfs}

We give some assumptions and definitions related to LTFs. Let $f$ be an LTF as in eqn.\ (\ref{eqn:ltf}). In general, the weights $\{w_i\}$ may be negative, and hence $f$ may not be monotone but only {\em locally monotone} (or {\em unate}). However, for the purposes of understanding query and communication complexity, it suffices to assume that the weights are indeed positive, as this may be simulated by local complementation of the individual bits. We also assume that the weights are given in non-increasing order, i.e.\ $w_1 \ge w_2 \ge \cdots \ge w_n$. Define $m_j$, where $j=0$ or $j=1$, as
\[ m_j = \min_{z,f(z)=j} \left|\sum_{i=1}^n w_i z_i - \theta \right|, \]
and define the {\em margin} of $f$ as $m = \min\,\{m_0,m_1\}$. We assume that there is no $x$ such that $\sum_{i=1}^n w_i x_i = \theta$, so the margin is strictly positive.


\subsection{Statement of results}
\label{sec:results}

Now we are equipped with definitions, the main results that we obtain can be stated concisely as follows.

\begin{itemize}
\item Section \ref{sec:detoneway}: If $g$ is an XOR function, then $D^{cc,1}(g) = Q_E^{cc,1}(g) = \dim f$.
\item Section \ref{sec:onewaysep}: For any positive integer $m$, there is an XOR function $g$ such that $D^{cc}(g) = O(m)$, but $Q_2^1(g) = \Omega(2^m)$.
\item Section \ref{sec:paritytrees}: For any XOR function $g$, $D^{cc}(g) = O(Q_E^{cc}(g))$, if the following conjecture holds: For any boolean function $f$, there exists a subset $T \subseteq [n]$ such that $|\supp(\hat{f}) \cap \supp(\hat{f}^{\Delta T})| \ge K \|\hat{f}\|_0$, for some constant $0 < K < 1$.
\item Section \ref{sec:randomised}: Let $g(x,y) = f(x \oplus y)$ be an XOR function. Then $R^{\|,pub}(g) = O(\|\hat{f}\|_1^2)$, and also $R^{\|,pub}(g) = O(\log(2^{n-1}(1-\|\hat{f}\|_\infty)))$. The former result is a special case of a theorem of Grolmusz \cite{grolmusz97}; we give a simplified proof.
\item Section \ref{sec:mono}: Let $g(x,y) = f(x \oplus y)$ be an XOR function. If $f$ is monotone, then $D^{cc}(g) = O(Q_E^{cc}(g)^2)$. If $f$ is an LTF and depends on $n$ bits, then $Q_E^{cc}(g) = \Omega(n)$.
\item Section \ref{sec:ltfprotocol}: Let $g(x,y) = f(x \oplus y)$ be an XOR function where $f$ is an LTF with margin $m$ and threshold $\theta$. Then $R^{\|,pub}(g) = O((\theta/m)^2)$.
\end{itemize}

We now turn to proving these results.


\section{Communication complexity of general XOR functions}


\subsection{Deterministic one-way communication complexity}
\label{sec:detoneway}

We begin by noting that the deterministic one-way communication complexity of {\em all} XOR functions has a simple characterisation.

\begin{prop}
Let $g(x,y) = f(x \oplus y)$ be an XOR function. Then
\[ D^{cc,1}(g) = Q_E^{cc,1}(g) = \dim f. \]
\end{prop}

\begin{proof}
It is well-known \cite{kushilevitz97} that $D^{cc,1}(g) = \lceil \log_2 \operatorname{nrows}(g) \rceil$, where $\operatorname{nrows}(g)$ denotes the number of distinct rows in the communication matrix of $g$, and Klauck showed that the same is true for deterministic one-way quantum communication \cite{klauck00}. Now it holds that
\beas \operatorname{nrows}(g) &=& \sum_{x \in \{0,1\}^n} \frac{1}{|\{y:f^{\oplus x} = f^{\oplus y}\}|} = \sum_{x \in \{0,1\}^n} \frac{1}{|\{y:f^{\oplus (x \oplus y)} = f\}|}\\
&=& \frac{2^n}{|\{y:f^{\oplus y} = f\}|} = \frac{2^n}{|\{y: \ip{y}{s} = 0\;\forall s \in \supp \hat{f}\}|}\\
&=& 2^{\dim f},
\eeas
where in the penultimate equality we use the fact (which follows easily from Fourier duality) that $f = f^{\oplus y}$ if and only if the function $\chi_y \cdot \hat{f} = \hat{f}$. This implies that there is no $s \in \supp \hat{f}$ such that $\ip{y}{s} = 1$, where the inner product is taken over $\F_2^n$.
\end{proof}


\subsection{Separation between one-way and two-way communication complexity}
\label{sec:onewaysep}

We now establish that there can be an exponential gap between the one-way (quantum, bounded-error) and two-way (classical, deterministic) communication complexity of XOR functions\footnote{Note that this is a {\em stronger} separation than between quantum and randomised communication complexity.}, using a VC-dimension argument. The VC-dimension of a matrix $M$, $\vcdim(M)$, is the largest $k$ such that there exists a $2^k \times k$ submatrix $M'$ of $M$ whose rows are all distinct. It was previously shown by Klauck \cite{klauck01} that VC-dimension gives a lower bound on bounded-error quantum communication complexity:

\begin{thm}[Klauck \cite{klauck01}]
\label{thm:vcbound}
Let $M$ be the communication matrix of some function $f$. Then $Q_2^1(f) = \Omega(\vcdim(M))$.
\end{thm}

We have the following proposition.

\begin{prop}
\label{prop:onewaysep}
Let $x$ be an $(m + 2^m)$-bit string divided into an $m$-bit ``address'' register $a$, and a $2^m$-bit ``data'' register $d$. Let $f(x)$ be the addressing function, which returns the data bit at a given address: $f(x) = d_a$. Finally, let $g$ be the XOR function $g(x,y) = f(x \oplus y)$. Then $D^{cc}(g) = O(m)$, but $Q_2^1(g) = \Omega(2^m)$.
\end{prop}

\begin{proof}
For the upper bound, note that $D(f) = m + 1$: a decision tree for $f$ can just evaluate the $m$ address bits, followed by the one relevant data bit. For the lower bound, we will show that $\vcdim(M) \ge 2^m$, with the result following from Theorem \ref{thm:vcbound}. Let $S_x$ be the set $\{(a,0^{2^m})\}$ for $a \in \{0,1\}^m$, and let $S_y$ be the set $\{(0^m,d)\}$ for $d \in \{0,1\}^{2^m}$. For all pairs of $2^m$-bit strings $d \neq d'$, there exists an $a$ such that $d_a \neq d'_a$. Thus, for all $y \neq y' \in S_y$, there is an $x \in S_x$ such that $f(x \oplus y) \neq f(x \oplus y')$, implying that $\vcdim(M) \ge 2^m$.
\end{proof}

Many of the most efficient known communication protocols for XOR functions require only one-way communication \cite{huang06,shi09}, and indeed it was left as an open question in \cite{shi09} whether all {\em symmetric} functions could be computed optimally using a one-way protocol. The above proposition implies that this cannot be true in a more general setting.


\subsection{Parity decision trees and Fourier spectra}
\label{sec:paritytrees}

We turn to the question of finding classical upper bounds, and quantum lower bounds, on the two-way deterministic communication complexity of XOR functions. This is where Fourier analysis becomes very useful, in particular because of the following natural observation, which appears to have first been written down by Shi and Zhang \cite{shi09}.

\begin{obs}
Let $g(x,y) = f(x \oplus y)$ be an XOR function. Then $D^{cc}(g) \ge \log_2 \|\hat{f}\|_0$ and $Q_E^{cc}(g) \ge \frac{1}{2}\log_2 \|\hat{f}\|_0$.
\end{obs}

\begin{proof}
Assume $f$ is a function on $n$ bits, and let $M$ be the communication matrix of $g$. Then it is easy to see that $M$ is diagonalised by the Fourier transform over $\Z_2^n$, and the eigenvalues of $M$ are given by $f$'s Fourier coefficients, scaled appropriately. Indeed, letting $F$ denote the matrix of this Fourier transform in the usual basis indexed by $n$-bit strings, $F_{xy} = (-1)^{\ip{x}{y}}$ (with the inner product being taken over $\F_2^n$), we have
\beas
\frac{1}{2^n} (F M F)_{xy} &=& \frac{1}{2^n} \sum_{u,v \in \{0,1\}^n} F_{xu} M_{uv} F_{vy} = \frac{1}{2^n} \sum_{u,v \in \{0,1\}^n} (-1)^{\ip{x}{u} + \ip{v}{y}} f(u \oplus v)\\
&=& \frac{1}{2^n} \sum_{w \in \{0,1\}^n} f(w) \sum_{u \in \{0,1\}^n} (-1)^{\ip{x}{u} + \ip{(w \oplus u)}{y}}
= \sum_{w \in \{0,1\}^n} f(w) (-1)^{\ip{w}{y}} \delta_{xy},
\eeas
which is equal to $2^n \hat{f}(x)$ if $x=y$, and 0 otherwise. So the rank of $M$ is equal to $\|\hat{f}\|_0$. The observation now follows from known results lower bounding the classical \cite{mehlhorn82} and quantum \cite{buhrman01a,nielsen98} communication complexity of a function by the log of the rank of its communication matrix.
\end{proof}

In the other direction, the following observation gives a natural way of finding {\em upper} bounds on the deterministic communication complexity of XOR functions.

\begin{obs}
\label{obs:paritytrees}
Let $g(x,y) = f(x \oplus y)$ be an XOR function. Then $D^{cc}(g) \le 2 D^\oplus(f)$.
\end{obs}

\begin{proof}
Given a parity decision tree for $f$ that uses at most $D^\oplus(f)$ queries on any input, a communication protocol for $g$ can be obtained as follows. Each query to a subset $S$ of the bits of the string $x \oplus y$ can be simulated by Alice sending the parity $\bigoplus_{i \in S} x_i$ to Bob, who reciprocates by sending her $\bigoplus_{i \in S} y_i$. This clearly enables each of them to compute $\bigoplus_{i \in S} (x_i \oplus y_i)$.
\end{proof}

Therefore, it would suffice to prove the following conjecture to show that quantum and classical communication complexity of XOR functions is polynomially related.

\begin{con}
\label{con:upper}
Let $f:\{0,1\}^n \rightarrow \{1,-1\}$ be a boolean function. Then
\[ D^\oplus(f) = O(\operatorname{polylog}(\|\hat{f}\|_0)). \]
\end{con}

It appears to be fairly difficult to reason about parity decision trees. We now give a conjecture which is merely about the structure of the Fourier spectrum of boolean functions and which, if true, would imply Conjecture \ref{con:upper}.

\begin{con}
\label{con:upper2}
Let $f:\{0,1\}^n \rightarrow \{1,-1\}$ be a boolean function. Then there exist universal constants $C$, $0 < K < 1$ such that, if $\|\hat{f}\|_0 > C$, there exists a subset $T \subseteq [n]$ such that $|\supp(\hat{f}) \cap \supp(\hat{f}^{\Delta T})| \ge K \|\hat{f}\|_0$.
\end{con}

In order to show that Conjecture \ref{con:upper2} does indeed imply Conjecture \ref{con:upper}, we will need the following lemma.

\begin{lem}
\label{lem:parityquery}
Let $f:\{0,1\}^n \rightarrow \R$ be some function on the boolean cube, and let $T\subseteq [n]$ be arbitrary. Define the function $g$ by
\[ g(x) = \left\{ \begin{array}{ll}f(x)&[\chi_T(x)=r]\\f(x\oplus t)&[\chi_T(x)=-r], \end{array} \right. \]
for some $t$ such that $\chi_T(t)=-1$, and some $r = \pm 1$. Then $g(x)=f(x)$ wherever $\chi_T(x)=r$, and for all $S$, $\hat{g}(S) = \frac{1}{2}((1+\chi_S(t)))(\hat{f}(S) + r \hat{f}(S \Delta T))$. In particular, for all $S$ such that $\chi_S(t)=-1$, $\hat{g}(S)=0$.
\end{lem}

\begin{proof}
The fact that $g(x)=f(x)$ wherever $\chi_T(x)=r$ is immediate; we now prove the second claim. We have
\begin{eqnarray*}
\hat{g}(S) &=& \frac{1}{2^n}\left(\sum_{x,\chi_T(x)=r} f(x) \chi_S(x) + \sum_{x,\chi_T(x)=-r} f(x+t) \chi_S(x)\right)\\
&=& \frac{1}{2\cdot 2^n}\left(\sum_{x \in \{0,1\}^n} (1+r \chi_T(x))f(x) \chi_S(x) + \sum_{x \in \{0,1\}^n} (1-r \chi_T(x))f(x+t) \chi_S(x) \right)\\
&=& \frac{1}{2\cdot 2^n}\left(\sum_{x \in \{0,1\}^n} (1+r \chi_T(x))f(x) \chi_S(x) + \sum_{x \in \{0,1\}^n} \chi_S(t)(1+r \chi_T(x))f(x) \chi_S(x)\right)\\
&=& \frac{(1+\chi_S(t))}{2 \cdot 2^n} \sum_{x \in \{0,1\}^n} (1+r \chi_T(x))f(x) \chi_S(x)\\
&=& \frac{(1+\chi_S(t))}{2} \left( \hat{f}(S) + r \hat{f}(S \Delta T) \right),
\end{eqnarray*}
which is clearly zero wherever $\chi_S(t)=-1$.
\end{proof}

Now consider an algorithm which attempts to evaluate $f(x)$ for some unknown input $x$ by making a query to the parity of the bits in a subset $T \subseteq [n]$, which is equivalent to querying the function $\chi_T(x)$. Given the knowledge that $\chi_T(x) = r$, for $r = \pm 1$, in order to evaluate $f(x)$, it suffices to evaluate $g(x)$ for any function $g$ of our choice, as long as $g(x) = f(x)$ wherever $\chi_T(x) = r$. That is, we can replace $f$ with $g$.

If we pick $g$ according to the procedure of Lemma \ref{lem:parityquery}, then as $\chi_T(t)=-1$, for each $S$ either $\hat{g}(S) = 0$, or $\hat{g}(S \Delta T) = 0$. This implies that whatever the value of $r$, the number of nonzero Fourier coefficients of $g$ is upper bounded by half of the number of subsets $S$ where either $\hat{f}(S) \neq 0$ or $\hat{f}(S \Delta T) \neq 0$; this quantity can be written down concisely as
\[ \frac{1}{2}\left|\supp(\hat{f}) \cup \supp(\hat{f}^{\Delta T})\right| = \|\hat{f}\|_0 - \frac{1}{2} \left|\supp(\hat{f}) \cap \supp(\hat{f}^{\Delta T})\right|. \]
%
So, if there exists a subset $T$ such that $|\supp(\hat{f}) \cap \supp(\hat{f}^{\Delta T})| \ge K \|\hat{f}\|_0$, for some constant $0 < K < 1$, then $\|\hat{g}\|_0$ will be at most a constant fraction of $\|\hat{f}\|_0$. If there exists such a subset for {\em all} boolean functions, then after repeating this procedure $O(\log \|\hat{f}\|_0)$ times (querying the parity of the bits in the best subset each time), $f$ would be reduced to a constant function. This would thus imply that $D^{\oplus}(f) = O(\log \|\hat{f}\|_0)$.


\section{Randomised protocols for XOR functions}
\label{sec:randomised}

In this section we discuss randomised classical protocols for computing general XOR functions. The first protocol we give is efficient for functions whose spectral norm is low\footnote{This is a special case of a result of Grolmusz \cite{grolmusz97}; we give a simplified proof.}, while the second is efficient for functions which are close to a parity function on some subset of the bits. These protocols can be seen as two different generalisations of a protocol for computing the equality function ($g(x,y) = 1 \Leftrightarrow x=y$), which satisfies both of these conditions. We give a third (!)\ generalisation of this protocol in Section \ref{sec:ltfprotocol}.

\begin{prop}[Grolmusz \cite{grolmusz97}]
\label{prop:spectral}
Let $g(x,y) = f(x \oplus y)$ be an XOR function with $f:\{0,1\}^n \rightarrow \{1,-1\}$. Then $R^{\|,pub}(g) = O(\|\hat{f}\|_1^2)$.
\end{prop}

\begin{proof}
We give a variant of a protocol of Kremer, Nisan and Ron \cite{kremer99} for computing the inner product of two vectors, which will achieve the specified complexity. Using their shared randomness, Alice and Bob pick $k$ subsets $\{S_i\}$ from the family of subsets of $[n]$, for some $k$ to be determined, where the set $S$ is picked with probability $|\hat{f}(S)|/\|\hat{f}\|_1$. For each subset $S_i$, Alice sends the referee the bit $\chi_{S_i}(x)$, and Bob sends the referee the bit $\chi_{S_i}(y)$. The referee uses these $k$ bits to compute
\[ \sum_{i=1}^k \chi_{S_i}(x) \chi_{S_i}(y) \sgn(\hat{f}(S_i)) = \sum_{i=1}^k \chi_{S_i}(x\oplus y) \sgn(\hat{f}(S_i)), \]
and outputs 1 if the result is positive, and $-1$ if negative. To see correctness of this protocol, note that for each $i$, $\chi_{S_i}(x\oplus y) \sgn(\hat{f}(S_i))$ is a sample from a random variable whose expectation is
\[ \frac{1}{\|\hat{f}\|_1} \sum_{S \subseteq [n]} \chi_S(x\oplus y) \hat{f}(S) = \frac{f(x \oplus y)}{\|\hat{f}\|_1}. \]
Standard Chernoff bound arguments thus give that the number of samples $k$ required to determine whether $f(x \oplus y)>0$, with a constant probability of success, is $O(\|\hat{f}\|_1^2)$.
\end{proof}

One can use the previous example of the addressing function to show that the above protocol is close to optimal in terms of its dependence on the spectral norm, even among all one-way quantum protocols. Indeed, the addressing function with an $m$-bit address register has spectral norm $2^m$, and by Proposition \ref{prop:onewaysep} has one-way quantum communication complexity $\Omega(2^m)$.

The second protocol rests on the following lemma.

\begin{lem}
\label{lem:closefn}
Let $f:\{0,1\}^n \rightarrow \{1,-1\}$ and $\tilde{f}:\{0,1\}^n \rightarrow \{1,-1\}$ be boolean functions that disagree on at most $m$ inputs, and let $g(x,y) = f(x \oplus y)$ and $\tilde{g}(x,y) = \tilde{f}(x \oplus y)$ be the corresponding XOR functions. Then $R^{\|,pub}(\tilde{g}) \le R^{\|,pub}(g) + O(\log m)$.
\end{lem}

\begin{proof}
Let $S$ be the set of inputs $z$ such that $f(z) \neq \tilde{f}(z)$. We give a protocol in the SMP model with shared randomness that determines whether $(x \oplus y) \in S$, using $O(\log |S|)$ bits of communication. This clearly implies the lemma: to get a protocol for $\tilde{g}$, it suffices to carry out the protocol for $g$, then check whether $(x \oplus y) \in S$, and if so, negate the result. In order to do this check, we use a simple generalisation of a well-known protocol for testing equality \cite{kushilevitz97}, which was also used by Gavinsky, Kempe and de Wolf \cite{gavinsky04} in their protocol for computing the Hamming distance. We give it explicitly for completeness.

Using their shared randomness, Alice and Bob create $k$ $n$-bit strings $\{r_1,\dots,r_k\}$, for some $k$ to be determined. Alice sends the referee the $k$-bit string $a = (\ip{x}{r_1},\dots,\ip{x}{r_k})$ that lists their inner products with $x$ over $\F_2$, and Bob does the same with the string $b = (\ip{y}{r_1},\dots,\ip{y}{r_k})$. The referee outputs 1 if there is some $z \in S$ such that $a_i \oplus b_i = \ip{z}{r_i}$ for all $i$, and otherwise outputs $-1$. We have
\[ \Pr[a_i \oplus b_i = \ip{z}{r_i}] = \Pr[\ip{x \oplus y}{r_i} = \ip{z}{r_i}], \]
which will equal 1 if $x \oplus y = z$, and $1/2$ otherwise. Thus the probability, for any given $z \in S$ with $x \oplus y \neq z$, that the referee incorrectly outputs 1 is $1/2^k$. Using a union bound over all $z \in S$, it suffices to take $k = O(\log |S|)$ to achieve a constant probability of success.
\end{proof}

Note that the above lemma still holds for stronger models of communication (e.g.\ $R_2^{cc}$, $R^1$), and that a similar result does not apparently hold for the communication complexity of general functions. It is now straightforward to see the following proposition.

\begin{prop}
Let $g(x,y) = f(x \oplus y)$ be an XOR function with $f:\{0,1\}^n \rightarrow \{1,-1\}$. Assume that there is some parity function $\chi_T$ such that $f$ disagrees with $\chi_T$ on $m$ inputs. Then $R^{\|,pub}(g) = O(\log m)$. In other words,
\[ R^{\|,pub}(g) = O(\log(2^{n-1}(1-\|\hat{f}\|_\infty))).\]
As a special case, if $f$ takes the value 1 (or the value $-1$) on at most $m$ inputs, $R^{\|,pub}(g) = O(\log m)$.
\end{prop}

\begin{proof}
It is clear that any function $g(x,y) = \chi_T(x \oplus y)$, with $T$ nonempty, has $R^{\|,pub}(g) = 2$ (by a protocol where Alice and Bob each send the referee the parity of the bits of their inputs in the set $T$). The result follows from Lemma \ref{lem:closefn}.
%
%
\end{proof}


\section{Communication complexity of monotone functions}
\label{sec:mono}

We now show that the two-way deterministic communication complexity of {\em monotone} XOR functions is almost determined by the rank. We will need the following lemma relating sensitivity and degree over $\F_2$; the proof is essentially the same as a previously known result relating sensitivity and degree over $\R$ \cite{buhrman02}.

\begin{lem}
\label{lem:sen}
Let $f:\{0,1\}^n \rightarrow \{0,1\}$ be a monotone boolean function. Then $s(f) \le \deg_2(f)$.
\end{lem}

\begin{proof}
It is well known (see \cite[Lemma 3]{bernasconi99}, for example) that the degree of $f$ over $\F_2$ is precisely the size of the largest subfunction of $f$ that takes the value 1 on an odd number of inputs. Now consider a point $x$ that achieves maximal sensitivity, i.e.\ $f(y) \neq f(x)$ for exactly $s(f)$ neighbours $y$ of $x$. Assume wlog $f(x)=1$. Now, by the monotonicity of $f$, all the points $z$ in the subcube traced out by $x$ and all the neighbours $y$ must have $f(z)=0$ (of the points in this subcube, $x$ must have maximal Hamming weight; for each $y$ neighbouring $x$, $f(y)=0$; and all other points in this subcube must have lower Hamming weight). So $f$ takes the value 1 on exactly one point in this dimension $s(f)$ subcube, so $\deg_2(f) \ge s(f)$.
\end{proof}

\begin{prop}
\label{prop:monotone}
Let $f:\{0,1\}^n \rightarrow \{0,1\}$ be a monotone boolean function. Define $g(x,y) = f(x \oplus y)$. Then $D^{cc}(g) \le 4(\log_2 \|\hat{f}\|_0)^2 = 4(\log_2 \rk g)^2$.
\end{prop}

\begin{proof}
The result follows from
\[ D^{cc}(g) \le 2D(f) \le 4s(f)^2 \le 4\deg_2(f)^2 \le 4(\log_2 \|\hat{f}\|_0)^2. \]
The inequalities are proven in order, as follows. For the first, if Alice and Bob have a decision tree for $f$, they can use it to compute $g$ with only an overhead of a factor of 2 \cite{kushilevitz97}. The second is proven as Corollary 5 of \cite{buhrman02}, while the third inequality follows from Lemma \ref{lem:sen}. The fourth is Lemma 3 of \cite{bernasconi99} (or see \cite[eqn.\ (2)]{gopalan09}).

\end{proof}

This proposition immediately implies the following corollary.

\begin{cor}
Let $f:\{0,1\}^n \rightarrow \{0,1\}$ be a monotone boolean function. Define $g(x,y) = f(x \oplus y)$. Then $D^{cc}(g) \le 16\;Q_E^{cc}(g)^2$.
\end{cor}


\subsection{Lower bounds on communication complexity of LTFs}

We turn to a class of XOR functions that is more specialised still: those based on linear threshold functions. We will see that the deterministic communication complexity of these functions is always $\Omega(n)$. We will need the following lemma, which does not appear to have been noted previously in the literature.

\begin{lem}
\label{lem:ltfsen}
Let $f$ be an LTF that depends on $n$ bits. Then $s(f) \ge \lceil (n+1)/2 \rceil$, and this result is best possible.
\end{lem}

\begin{proof}
Write the weights in non-increasing order, $w_1 \ge \dots \ge w_n$. Then, as $f$ depends on all $n$ variables, there exists an assignment to the bits $x_1,\dots,x_{n-1}$ such that
\[ \sum_{i=1}^{n-1} w_i x_i + w_n > \theta, \]
but
\[ \sum_{i=1}^{n-1} w_i x_i < \theta. \]
Call this assignment $(z_1,\dots,z_{n-1})$. As $w_n$ is the smallest of the weights, flipping any of the bits of the string $z^0 = (z_1,\dots,z_{n-1},0)$ from 0 to 1 will change the value of $f$, as will flipping any of the bits of the string $z^1 = (z_1,\dots,z_{n-1},1)$ from 1 to 0. Thus $s(f)$ is lower bounded by the maximum of $\{n-|z^0|,|z^1|\}$, which is at least $\lceil (n+1)/2 \rceil$. The {\sc Majority} function has sensitivity $\lceil (n+1)/2 \rceil$ and demonstrates that this result is best possible.
\end{proof}

\begin{prop}
Let $f$ be an LTF that depends on $n$ bits, and set $g(x,y) = f(x \oplus y)$. Then $D^{cc}(g) \ge \lceil (n+1)/2 \rceil$ and $Q_E^{cc}(f) \ge \lceil (n+1)/4 \rceil$.
\end{prop}

\begin{proof}
In the proof of Proposition \ref{prop:monotone} it was shown that, if $f$ is monotone, $\log_2 \rk(g) \ge s(f)$. The present proposition now follows from Lemma \ref{lem:ltfsen} and known results lower bounding classical \cite{mehlhorn82} and quantum \cite{buhrman01a,nielsen98} communication complexity by the log of the rank of $g$.
\end{proof}


\subsection{Upper bounds on communication complexity of LTFs}
\label{sec:ltfprotocol}

The final result of this paper is an upper bound on the randomised classical communication complexity of LTFs, derived by giving an explicit protocol for such functions in the SMP model with shared randomness. Formally, we have the following result.
\begin{prop}
Let $g(x,y) = f(x \oplus y)$, where $f$ is an LTF with threshold $\theta$ and margin $m$. Then $R^{\|,pub}(g) = O((\theta/m)^2)$.
\end{prop}

Our protocol can be seen as a generalisation of Yao's protocol for the Hamming distance function \cite{yao03}, which in turn can be understood as a generalisation of the well-known constant-communication protocol for computing equality of two bit strings. It proceeds as follows.

\begin{enumerate}
\item Alice and Bob use their shared randomness to generate $k = O((\theta/m)^2)$ $n$-bit strings $r_1, \dots, r_k$, where the $i$'th bit of each string $r_j$ is equal to 1 with probability $p_i$, for some probabilities $\{p_i\}$ which will be determined later.

\item For each $j$, Alice and Bob each compute the bits $a_j = \ip{r_j}{x}$ and $b_j = \ip{r_j}{y}$ (respectively), where the inner product is taken over $\F_2^n$, and each send the resulting $k$ bits to the referee.

\item The referee computes $s = \frac{1}{k}\sum_{j=1}^k (a_j \oplus b_j)$ and outputs 1 if
\[ s > \frac{1}{2}\left( 1 - \frac{1}{2} \left((1-1/\theta)^{\theta - m_0} + (1-1/\theta)^{\theta + m_1} \right) \right). \]
where $m_0$, $m_1$ are defined as in Section \ref{sec:ltfs}, and we assume that $m_0$, $m_1$, and $\theta$ are all greater than 1, rescaling if necessary.
\end{enumerate}

We now prove that there is a choice of $\{p_i\}$ such that this protocol succeeds with constant probability. We will need the following lemma.

\begin{lem}
\label{lem:parityprob}
Let $x$ be an arbitrary $n$-bit string, and let $r$ be a randomly generated $n$-bit string such that $\Pr[r_i=1] = p_i$ for some $\{p_i\}$. Then
\[ \Pr_r[\ip{r}{x} = 1] = \frac{1}{2} \left(1 - \prod_{i=1}^n \left(1 - 2\,p_i x_i\right) \right). \]
\end{lem}

\begin{proof}
For $1 \le k \le n$, define $Q_k = \Pr_r[ \bigoplus_{i=1}^k r_i x_i = 1]$. Then, for $2 \le k \le n$,
\beas
Q_k &=& (1 - \Pr_r[ \bigoplus_{i=1}^{k-1} r_i x_i = 1])\Pr_r[r_k x_k = 1] + \Pr_r[ \bigoplus_{i=1}^{k-1} r_i x_i = 1](1 - \Pr_r[r_k x_k = 1])\\
&=& Q_{k-1}(1 - 2 p_k x_k) + p_k x_k,
\eeas
and also $Q_n = \Pr_r[\ip{r}{x} = 1]$. Now the lemma follows by induction on $k$, noting that the base case
\[ Q_1 = p_1 x_1 = \frac{1}{2} \left( 1 - \prod_{i=1}^1 (1 - 2 p_i x_i) \right). \]
\end{proof}

Now the central idea behind our approach is as follows. Consider the string $z = x \oplus y$. The referee needs to output 1 if $\sum_{i=1}^n w_i z_i > \theta$. He does not know $\sum_{i=1}^n w_i z_i$, but if we pick $p_i$ to be small and proportional to $w_i$, the quantity
\[ \prod_{i=1}^n \left(1 - 2\,p_i z_i\right), \]
which the referee {\em can} estimate using Lemma \ref{lem:parityprob}, should give an estimate of $\sum_{i=1}^n w_i z_i$, as the first order terms are proportional to this sum. We will not in fact quite do this, but will do something easier to analyse. If we pick
\[ p_i = \frac{1}{2}\left(1- (1-2\alpha)^{w_i} \right), \]
for some constant $0 \le \alpha \le 1$ to be determined, we get
\be \label{eqn:prob} \Pr_r[\ip{r}{z}=1] = \frac{1}{2} \left(1 - \prod_{i=1}^n (1-2\alpha)^{w_i z_i} \right) = \frac{1}{2} \left(1 - (1-2\alpha)^{\sum_{i=1}^n w_i z_i} \right). \ee
Write $v = \sum_{i=1}^n w_i z_i$. Our task is now to choose a value for $\alpha$ that makes the two cases $v < \theta$, $v > \theta$ easy to distinguish. As the most difficult cases to distinguish will be when $v \approx \theta$, we achieve this by choosing $\alpha$ to maximise the absolute value of the derivative
\[ \frac{d}{dv} \frac{1}{2} \left(1 - (1-2\alpha)^v \right) = -\frac{1}{2} (1-2\alpha)^v \ln(1-2 \alpha), \]
evaluated at $v = \theta$. For $0 < \alpha < 1/2$ this derivative is positive, and we have
\[ \frac{d}{d \alpha}\left(-\frac{1}{2} (1-2\alpha)^\theta \ln(1-2 \alpha) \right) = (1-2\alpha)^\theta(1 + \theta \ln(1-2 \alpha)). \]
Setting this expression equal to 0 and solving for $\alpha$ gives
\[ \alpha = \frac{1}{2} \left( 1 - e^{-1/\theta} \right) \approx \frac{1}{2 \theta}. \]
Inserting this value for $\alpha$ into eqn. (\ref{eqn:prob}), we obtain
\[ \Pr_r[\ip{r}{z}=1] = \frac{1}{2} \left(1 - (1-1/\theta)^{\sum_{i=1}^n w_i z_i} \right). \]
Our problem has therefore been reduced to determining whether $\sum_{i=1}^n w_i z_i > \theta$, using samples from this distribution. The remainder of the proof is a standard Chernoff bound argument. Let $X$ denote the sum of $k$ i.i.d.\ random variables $X_i$, which take values in $\{0,1\}$, with $\Pr[X_i=1] = \mu$. Then the inequality
\[ \Pr[|X-k \mu|>\delta] < 2e^{-\delta^2/4k\mu} \]
holds, implying that one can distinguish two different distributions with means $\mu$, $\mu'$, where $|\mu - \mu'| \ge \epsilon$, with $O(1/\epsilon^2)$ samples from $X_i$.

Recall that $|\sum_{i=1}^n w_i z_i - \theta| \ge m$ for all $z$. Thus, for any $z$, $z'$ such that $f(z) \neq f(z')$, we have
\beas
|\Pr_r[\ip{r}{z}=1] - \Pr_r[\ip{r}{z'}=1]| &\ge& \frac{1}{2} \left((1-1/\theta)^{\theta - m} - (1-1/\theta)^{\theta + m} \right)\\
&=& \frac{1}{2}(1-1/\theta)^{\theta} \left((1-1/\theta)^{-m} - (1-1/\theta)^m \right)\\
&=& \Omega(m/\theta),
\eeas
which implies that it suffices for the referee to take $O((\theta/m)^2)$ samples from the distribution to determine whether $\sum_{i=1}^n w_i z_i > \theta$ with constant probability. The threshold value picked in the protocol is simply halfway between the two worst-case values of $z$.


\section{Conclusions}

We have presented a number of partial results on the communication complexity of XOR functions, but the initial question still remains: are the quantum and classical communication complexities of XOR functions polynomially related? We believe that the class of XOR functions is of particular interest in the context of communication complexity because of the connection to Fourier analysis of boolean functions, and remain hopeful that this conjecture is tractable. The little-studied classical model of parity decision tree complexity also appears to be of some interest in its own right; the connection with the ``width'' of the Fourier spectrum is an interesting contrast to the usual decision tree complexity, which is polynomially related to the ``height'' (degree) of the Fourier spectrum.

A final question: can the protocol of Section \ref{sec:ltfprotocol} be improved to use, for example, $O((\theta/m) \log (\theta/m))$ communication, in a similar way to Huang et al's protocol for the Hamming distance problem \cite{huang06}?


\section*{Acknowledgements}

AM was supported by the EC-FP6-STREP network QICS and an EPSRC Postdoctoral Research Fellowship, and would like to thank Aram Harrow and Rapha\"el Clifford for helpful comments on a previous version.



	
\end{document}